# The Behavior of the Intercalant AlCl$_4$ Anion during the Formation of Graphite Intercalation Compound: An X-ray Absorption Fine Structure Study

*Giorgia Greco,\* Giuseppe Antonio Elia,\* Yves Kayser, Burkhard Beckhoff, Marko Perestjuk, Simone Raoux, and Rober Hahn*

This work aims to study the insertion of AlCl$_4^-$ anion in the crystalline structure of oriented pyrolytic graphite (PG) at the point of view of the anion itself. The electronic and atomic structures of the anion at different intercalation stages are studied. In particular double-edge (bicolor) X-ray absorption spectroscopy at the Al and Cl *K*-edges is carried out, highlighting a contraction of the anion bonding at the highest intercalation degree obtained electrochemically (stage 3), while the electronic population changes for both the edges upon cycle.

## 1. Introduction

Understanding the processes behind the phase transitions, particularly the property of some solids to host different species, is of extreme interest. Indeed this phenomenon has a practical application in energy storage and conversion systems such as high-energy-density batteries and hydrogen storage, which are very important topics due to the need to store and convert energy from renewable sources.[1] The first step for material design and engineering to reinforce our energy storage capacity is a comprehensive structural characterization which is a challenging research activity.

In our previous studies, we have comprehensively characterized the graphite intercalation compound (GIC) formation in aluminum graphite dual-ion cell (AGDIC) by combining operando small- and wide-angle X-ray scattering[2] and tomography with X-ray diffraction (XRD).[3] A detailed electrochemical characterization of the system is reported in our previous studies.[2–4] We have directly observed the graphite structural deformation induced by the electrochemical intercalation process and its intermediate states, which result in two-phase transitions from graphite to GIC with 0.05 mole of anion intercalated for 1 mol of graphite or stage 3.[2,3] The intercalation stage is defined as the numbers of free graphene layers between two layers of intercalant species; stage 3 means 3 graphene layers between two layers of AlCl$_4^-$ anion. We also highlighted the differences

G. Greco,[+] M. Perestjuk,[++] S. Raoux
Helmholtz-Zentrum Berlin für Materialien und Energie GmbH
Hahn-Meitner-Platz 1, D-14109 Berlin, Germany
E-mail: giorgia.greco@uniroma1.it

The ORCID identification number(s) for the author(s) of this article can be found under https://doi.org/10.1002/pssa.202300776.

[+]Present address: Department of Fusion and Technology for Nuclear Safety and Security, ENEA Centro Ricerche Casaccia, Via Anguillarese 301, 00123 Rome, Italy

[++]Present address: Université de Lyon, Institut des Nanotechnologies de Lyon (INL) UMR CNRS 5270, Ecole Centrale Lyon, 69131 Ecully, France

[+++]Present address: Department of Applied Science and Technology (DISAT), Politecnico di Torino, Corso Duca degli Abruzzi 24, 10129 Torino, Italy

[++++]Present address: Max Planck Institute for Chemical Energy Conversion (MPI CEC), Stiftstr. 34-36, 45470 Mülheim an der Ruhr, Germany

G. Greco,
Chemistry Department
Sapienza University of Rome
P.le Aldo Moro 5, 00185 Roma, Italy

G. A. Elia,[+++] R. Hahn
Technical University of Berlin
Research Center of Microperipheric Technologies
Gustav-Meyer-Allee 25, D-13355 Berlin, Germany
E-mail: giuseppe.elia@polito.it

Y. Kayser,[++++] B. Beckhoff
Physikalisch Technische Bundesanstalt (PTB)
Abbestr. 2-12, 10587 Berlin, Germany

M. Perestjuk,
School of Engineering
RMIT University
Melbourne, VIC 3001, Australia

R. Hahn
Fraunhofer IZM Institut für Zuverlässigkeit und Mikrointegration
Gustav-Meyer-Allee 25, D-13355 Berlin, Germany

between two types of graphite, natural (NG) and pyrolytic (PG), showing how the difference in the elasticity of the crystalline structure and the mesoscopic porosity influences the reversibility of the intercalation process.[3,5] Nevertheless, some aspects, such as this system's limited capacity, are still unclear. Theoretically, each graphene layer could host a layer of intercalant species reaching stage 1, but there is an empirical limitation to stage 3.

In this article we study the behavior of $AlCl_4^-$ anions during the intercalation process in PG layers by X-ray absorption fine structure (XAFS) spectroscopy at both the Cl and Al absorption K-edges. This bicolor technique being selective to two elements gives a new insight into the intercalation process from the point of view of the intercalant species, clarifying some aspects that have so far remained questionable. The article is structured as follows.

The first part shows the electronic structure of the anion at Al and Cl K-edges for different intercalation stages. The second part focuses on quantifying the anion uptake at the different intercalation stages, while the third one aims to highlight the structural changes due to the strain and the compression of the graphite lattice on the anion.

## 2. Results and Discussion

### 2.1. Samples

The electrochemical processes of the AGDIC function are the following.

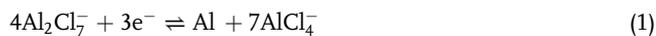
$$4Al_2Cl_7^- + 3e^- \rightleftharpoons Al + 7AlCl_4^- \tag{1}$$

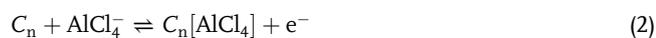
$$C_n + AlCl_4^- \rightleftharpoons C_n[AlCl_4] + e^- \tag{2}$$

During the charging process at the negative electrode, $Al_2Cl_7^-$ anions in the ionic liquid electrolyte react at the Al anode side forming $AlCl_4^-$ anions and Al metal. Concurrently at the positive electrode, $AlCl_4^-$ anions intercalate into the PG graphitic structure. **Figure 1** shows the typical voltage profile for this system. The arrows in the figure mark the state of charge selected for the ex situ investigation, that is, 25 Ch. (25 mAh g$^{-1}$ charged), 50 Ch. (50 mAh g$^{-1}$ charged), fully Ch. (fully charged), and the same for discharging. As already described in our previous work,[3] PG shows an irreversible capacity of about 30% due to partial trapping of a certain amount of anions in the PG structure. This phenomenon is most likely associated with the PG structure's low elasticity and poor porosity.[5] Ex situ XAFS spectroscopy at both the Al and Cl absorption K-edges experiments was carried out on different PG electrodes at different intercalation degrees[5] (see **Table 1**) to follow the intercalant behavior in the host at different states of charge. Moreover, samples obtained at the 5th, 500th, and 1000th cycles have been investigated to evaluate the long-term cycling effect. A more detailed description of sample preparation can be found in the Experimental Section.

### 2.2. XANES

In order to better understand the molecular composition of the intercalated species, first we compared the X-ray absorption near-edge spectra (XANES) jump (absorption discontinuity due to the ionization cross section at the K-edge) at the Al and Cl K-edges, as shown in **Figure 2** for the fully discharged sample. As already described in our previous work,[3] the electrodes expanded noticeably up to 300% from the initial thickness of ≈20 μm at the fully charged state (fully intercalated). Due to incorporated material, we could not collect at Al K-edge (1559.6 eV) in transmission mode, but only in fluorescence mode. For this reason, we reveal information of the evaluated molecule stoichiometry for a total of five samples (see Table 1). The XANES jump (absorption discontinuity) in transmission mode is proportional to the actual areal mass density (mass per unit area as integrated into the depth direction) of an element in the sample.[6,7] Thus, the $AlCl_4^-$-intercalated stoichiometry in PG could be obtained from the jump ratio between Al and Cl K-edges, knowing the cross section of the elements for the incident photon energies used around the respective ionization thresholds.[6] The results are shown in Table 1. The expected stoichiometry 4:1 Cl:Al atomic ratio is confirmed for all the samples except for the sample after 500 cycles, most likely associated with Cl accumulation upon cycling with the formation of chlorine species on the electrode surface consistent with the presence of $CCl_x$ and $AlCl_x$ species.[4] This phenomenon has also been evidenced in our previous study by X-ray photoemission spectroscopy (XPS).[4] As confirmed by the actual stoichiometry of the anion, we can calculate with high accuracy the mass uptake of the molecule in GIC by the formula

$$m_{AlCl_4^-} = N_{AlCl_4^-}(m_{at,Al} + 4\, m_{at,Cl}) \tag{3}$$

where $N_{AlCl_4^-}$ is the number of uptake molecules, which can be evaluated from XAS jump and the cross section of the element and[6] $m_{at,m}$ (where $m$ = Al, Cl) is the elemental atomic mass. Results in Table 1 are cross-checked at Al and Cl K-edges and are in good agreement with the estimation of the $AlCl_4^-$ uptake from the electrochemical results.[3]

Going forward, we want to study the transformations of the electronic structure upon cycling. **Figure 3** shows the normalized near-edge XANES spectra at the Al and Cl K-edges collected in fluorescence and transmission mode, respectively, for a set of electrodes in a subsequent intercalation stage. Differences in the intensity of the spectral features (labeled in Figure 3 as A–D for the Al K-edge data and A′–C′ for the Cl K-edge data) are clearly visible for both edges. The normalized spectra are compared with AlCl$_3$ powder as reference and with Al metal at the Al K-edge. Looking at Figure 3a (Al K-edge), an edge shift of $\Delta E = -1.32$ eV and strong decrease of the peak B–C of the signal of the electrodes compared to the AlCl$_3$ powder one is observed. At Cl K-edge from AlCl$_3$ to AlCl$_4^-$, a different behavior (Figure 3) is evident; while the white line increases (B′ peak), the peaks A′ and C′ slightly decrease. This effect is due to a change in the coordination number of the Al—Cl and Cl—Cl bonds from three- to fourfold and from two to threefold, respectively.[8] The oxidation state of Al and Cl is always +3 and −1 respectively, for both structures.

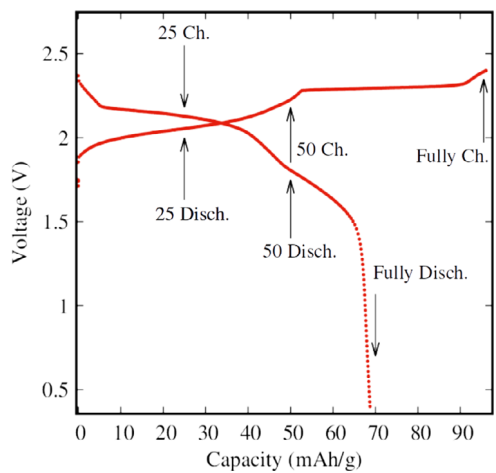

**Figure 1.** Voltage profile of the PG.

Table 1. In the columns, sample names and the related specific capacity in mAh g$^{-1}$, the intercalation stage number, the calculated atomic ratio between Al and Cl, and the calculated anion mass uptake in mg obtained from Al and Cl XAS jump ratio (absorption discontinuity due to the ionization cross section at the K-edge), respectively, are reported.

| Sample | Specific Capacity [mAh g$^{-1}$] | Stage | Al:Cl [atm] | AlCl$_4^-$ Mass from Al K-edge [mg] | AlCl$_4^-$ Mass from Cl K-edge [mg] |
|---|---|---|---|---|---|
| 1 cycle, 25 mAh g$^{-1}$ charged | ~25 | 6 | (4.2 ± 0.7):1 | 1.6 ± 0.3 | 1.7 ± 0.1 |
| 1 cycle, 50 mAh g$^{-1}$ charged | ~50 | 5, 4[a] | | – | 3.0 ± 0.1 |
| 1 cycle, fully charged | ~100 | 3 | | – | 3.4 ± 0.1 |
| 1 cycle, 25 mAh g$^{-1}$ discharged | ~75 | 3, 4[a] | | – | 3.6 ± 0.1 |
| 1 cycle, 50 mAh g$^{-1}$ discharged | ~50 | 4, 5[a] | (4.0 ± 0.7):1 | 3.0 ± 0.3 | 3.3 ± 0.1 |
| 1 cycle, fully discharged | ~30 | 6 | (4.0 ± 0.5):1 | 2.3 ± 0.3 | 2.3 ± 0.1 |
| 5 cycle, 25 mAh g$^{-1}$ charged | ~55 | 5 | (4.0 ± 0.4):1 | 3.0 ± 0.3 | 3.0 ± 0.1 |
| 5 cycles, 50 mAh g$^{-1}$ charged | ~80 | 4 | | – | 3.4 ± 0.1 |
| 5 cylcles, fully charged | ~100 | 3 | | – | 3.4 ± 0.1 |
| 5 cycles, 25 mAh g$^{-1}$ discharged | ~75 | 3, 4[a] | | 3.0 ± 0.3 | – |
| 5 cycles, 50 mAh g$^{-1}$ discharged | ~50 | 4, 5[a] | | – | 2.6 ± 0.1 |
| 5 cycles, fully discharged | ~30 | 6 | (3.5 ± 0.4):1 | 2.3 ± 0.3 | 2.0 ± 0.1 |
| 500 cycles, fully discharged | ~30 | 6 | (5.4 ± 0.9):1 | 1.6 ± 0.3 | 2.2 ± 0.1 |
| 1000 cycles, fully discharged | ~30 | 6 | | – | 2.1 ± 0.1 |

[a]Coexistence of two stages.

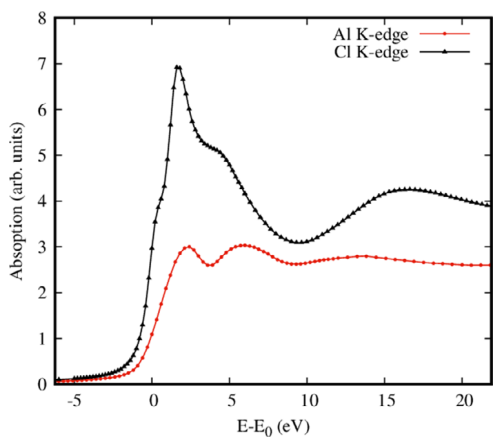

Figure 2. Comparison between near-edge XAS spectra at Al and Cl K-edge of one electrode fully discharged after the first cycle. $E_0$ is the edge energy.

Each peak of the XANES signal is associated with an electronic transition as reported in **Table 2** for both the Al and Cl K-edges.[8–11]

Figure 3a's inset compares normalized XANES signals at the Al K-edge of electrodes at subsequential intercalation stages (see also Figure 1 and Table 1). The Al K-edge signal evidences the presence of four main peaks, labeled A, B, C, D. The main peaks A and B decrease from 25 Ch. to 25 Disch. and again increase at Fully Disch. state. The Cl K-edge signal evidences the presence of four main peaks, labeled A′, B′, and C′. Similarly, the Cl K-edge peak B′ related to the main transition (see Table 1 and 2) decreases from 25 Ch. to Fully Ch. and increases again at Fully Disch. state. It has to be noted that the intensities observed for the Fully DisCh sample do not reach the initial state condition as expected,[4] due to the partial retention of AlCl$_4^-$ in the PG structure, as shown in Figure 1.

### 2.3. Double-Edge Analysis EXAFS

In order to comprehensively analyze the AlCl$_4^-$ atomic anion structure during the intercalation process, a double-edge analysis (Al and Cl K-edges) of the extended X-ray absorption fine structure (EXAFS) part of the XAFS data is presented. An advanced technique based on multiple scattering (MS) calculations of the absorption cross section in the framework of GNXAS program[12,13] was used. This method has already been applied to different functional materials for energy conversion and storage.[7,14,15] **Figure 4**a,c shows the best fit and its Fourier transform (FT) obtained for the Fully Disch. sample as an example. As shown in Figure 4, three bodies Al–Cl, Cl–Al, and Cl–Cl signal were used to fit the EXAFS signal for both Al and Cl K-edges. Figure 4b shows the theoretical structure of AlCl$_4^-$ calculated by the Avogadro[16] program. The fitting procedure was applied to all the electrodes at different intercalation stages, in **Table 3** the main results are shown. The first consideration is the significant difference between the AlCl$_3$ theoretical and intercalated one of Al—Cl/Cl—Al bond distance ($\Delta R_{\text{Al–Cl/Cl–Al}} \sim 0.07$ Å). Moreover the interatomic distance between Cl–Cl is shrunk with respect to theory ($\Delta R_{\text{Cl–Cl}} \sim 0.07$ Å). Upon cycling, the interatomic distances remain unchanged up to the fully charged state (stage 3). Here we have a clear contraction of $\Delta R_{\text{Al–Cl/Cl–Al}} \sim 0.01$ and 0.02 Å for the electrode fully charged for the first cycle and after five cycles, respectively (see Table 3). **Figure 5** compares the EXAFS signals and its FT of the 25 Ch. with the Fully Ch. after five cycles, for both edges. It is possible to observe a contraction of the interatomic distances in particular at the Al $K$-edge and a decrease in the FT intensities, showing an increase in the structural disorder. The EXAFS study shows a contraction of AlCl$_4^-$ bonding for both edges at the fully charged state, which corresponds to stage 3. The FT of the EXAFS signal also shows an intensity decrease at Cl and Al $K$-edges (Figure 5). This reveals an increase in the disorder of AlCl$_4^-$. This is probably due to the high level of strain of graphite structure measured in our previous work.[3] The anions first get a preferential orientation to occupy less space and then start to contract the bonding length up to a maximum level corresponding to stage 3.[2,3]

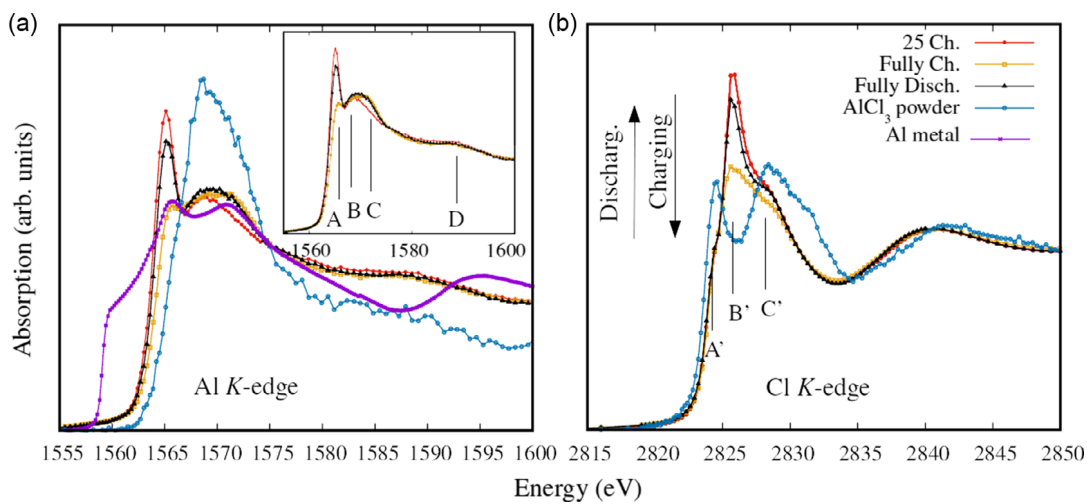

**Figure 3.** Normalized a) Al and b) Cl $K$-edge XANES collected in fluorescence (Al-$K$-edge) and transmission mode (Cl $K$-edge) of a set of electrodes subsequently charged and discharged compared to the signal obtained for AlCl$_3$ powder and Al metal.

**Table 2.** Al $K$-edge and Cl $K$-edge: Peaks label depicted in Figure 2, its position (energy) and the electronic transition related to.

| Electrodes | | |
|---|---|---|
| Edge $E_0 = 1563.46$ eV | | |
| Peak | Energy [eV] | Related transition |
| A | 1565.15 | 1 s → 3p ($t_2$) |
| B | 1568.80 | MS |
| C | 1576.36 | 1 s → 3d (e) |
| D | 1587.00 | 1 s → 3d ($t_2$) |
| AlCl$_3$ powder | | |
| Edge $E_0 = 1564.84$ eV | | |
| Peak | Energy [eV] | Related transition |
| A | 1568.80 | 1 s → 3p ($t_{1u}$) |
| Electrodes | | |
| Edge $E_0 = 2823.53$ eV | | |
| A' | 2824.40 | 1 s → 3d/3p |
| B' | 2825.73 | 1s → 4p |
| C' | 2828.20 | 1 s → higher states in Cl ion |
| AlCl$_3$ powder | | |
| Edge $E_0 = 2823.53$ eV | | |
| A' | 2824.40 | 1 s → 3d/3p |
| C' | 2828.20 | 1 s → higher states in Cl ion |

## 3. Conclusion

An XAFS study of the AlCl$_4^-$ anion intercalated in PG at different intercalation stages has been studied at Al and Cl $K$-edges. The XANES structure shows differences in the features and intensities of both edges revealing an active part of the Cl in the electronic exchange with PG upon cycling. The high level of strain in the PG structure causes an increase in the disorder level in the local structure detected at both edges. From the double-edges analysis, AlCl$_4^-$, the last intercalation stage 3 shrunk its interatomic distances probably due to the high level of strain in the PG structure.[3] The obtained information gives further insight into the mechanism of anion intercalation in graphite, thus potentially leading to a more rational design of cell components for high-performance batteries.

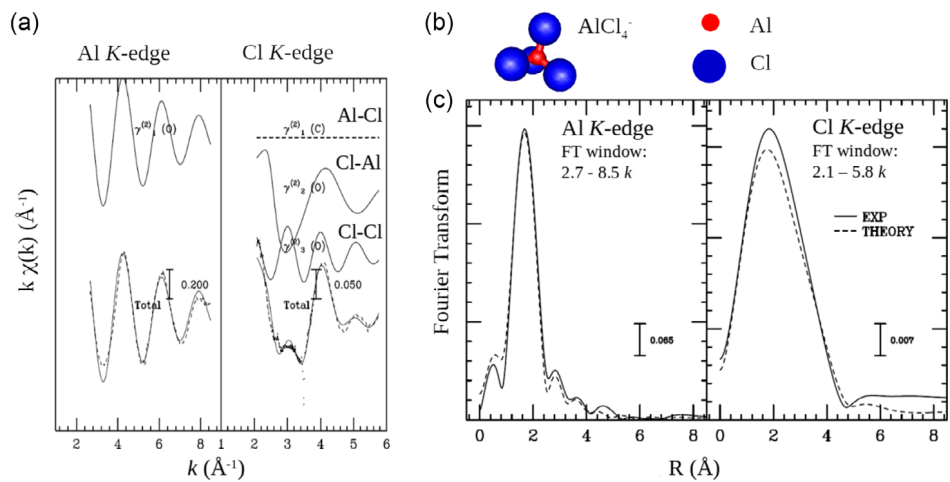

**Figure 4.** Double-edge best fit of the fully discharged electrode. EXAFS signal and its FT, a) and c) panel, respectively. The theoretical molecular structure of the $AlCl_4^-$ anion in panel (b) is shown.

**Table 3.** Best fit results of the electrodes at a different intercalation stage. Elect. (mid. stages) indicates the electrodes except for the fully charged states electrodes.

| Sample | $N_{Al-Cl/Cl-Al}$ | $R_{Al-Cl/Cl-Al}$ [Å] | $\sigma_{Al-Cl/Cl-Al}$ [Å$^2$] | $N_{Cl-Cl}$ | $R_{Cl-Cl}$ [Å] | $\sigma_{Cl-Cl}$ [Å$^2$] |
|---|---|---|---|---|---|---|
| $AlCl_4^-$ Theor. | 4/1 | 2.17 | – | 3 | 3.54 | – |
| Elect. (mid. stages) | " | 2.10 | 0.001/0.011 | " | 3.47 | 0.020 |
| Fully Ch. I | " | 2.09 | 0.010/0.010 | " | 3.46 | 0.024 |
| Fully Ch. V | " | 2.08 | 0.004/0.011 | " | 3.44 | 0.022 |
| Fully Disch. | " | 2.10 | 0.001/0.010 | " | 3.48 | 0.023 |

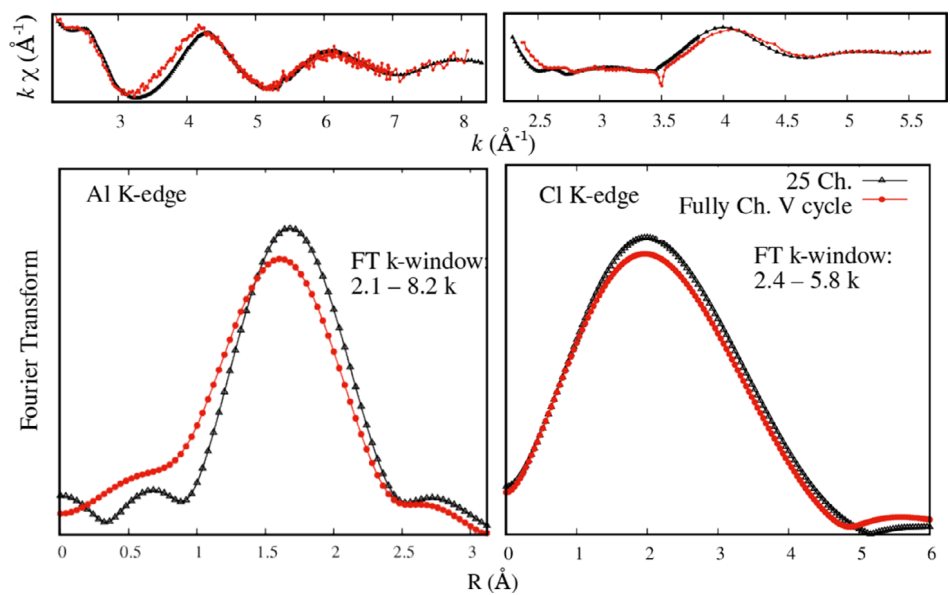

**Figure 5.** FT of the EXAFS signal at Al and Cl $K$-edges of the electrodes 25 mAh g$^{-1}$ charged I cycle compared to the fully charged 5th cycle.

## 4. Experimental Section

*Sample Preparation*: The electrolyte 1-ethyl-3-methylimidazolium chloride:aluminum trichloride EMIMCl:AlCl3 in a 1:1.5 mole ratio was provided by IOLITEC. The water content of the electrolyte was lower than 100 ppm. The electrochemical tests were performed using high-purity aluminum (Al 99.99% Alfa Aesar) as anode and PG with a thickness of 25 μm and a loading of 4.71 mg cm$^{-2}$ as the cathode material.[17,18] The electrochemical measurements were performed using Teflon Swagelok-type T cells. The cycling tests of Al/EMIMCl:AlCl3/PG cells were carried out applying increasing specific currents of 25 mA g$^{-1}$) in the voltage range 0.4–2.4 V. The upper cutoff was selected within the electrochemical stability of the electrolyte, evaluated by LSV to be 2.45 V versus Al quasireference.[4] All galvanostatic cycling tests were carried out at 25 °C in a thermostatic climatic chamber (with a possible deviation of ±1 °C), using a Maccor 4000 Battery Test System. Prior to the ex situ analyses, the studied electrodes were rinsed in order to remove residual electrolyte using dimethyl carbonate, and rigorous anhydrous and water content was detected by the Karl Fischer titration method. The preparation of the electrode was performed in the glovebox, with oxygen and water content below 1 ppm to avoid the electrode's degradation.

*Al K-Edge XAFS*: The XAFS were realized in the near-edge region of the Al K-edge (i.e., XANES measurements ranging from ≈1550 to 1600 eV for Al) and the extended regime (i.e., EXAFS measurements ranging from ≈1570 to 1840 eV for Al) by varying the incident photon energy from 1490 to 1837.5 eV in variable-energy steps at the plane-grating monochromator beamline of PTB for undulator radiation.[18] In the near-edge X-ray absorption fine structure (NEXAFS) range, a constant increment in the incident photon energy of 0.3 eV was selected, in the EXAFS regime, the incident photon energy was varied such that the change in k-space was constant (0.05 Å$^{-1}$). At each incident photon energy, the X-ray spectrum recorded by the silicon drift detector (SDD) was integrated over 20 s and the signal of the diode downstream of the sample was averaged over the same time interval. The samples were oriented at 45° with respect and with the normal vector to the sample surface plane contained within the polarization plane of the incident X-ray photons. The SDD was positioned in the polarization plane and perpendicular to the propagation direction of the linearly polarized incident X-ray beam in order to minimize scattered radiation.

*Cl K-Edge XAFS*: The XAFS measurements on Cl were carried out at PTB's four-crystal-monochromator beamline.[17] It provides monochromatized radiation between 1.75 and 10.5 keV by means of either four InSb(111) or Si(111) crystals. The use of four monochromator crystals allowed for the provision of X-ray radiation with a high spectral resolving power of 10$^4$.[17] Furthermore, the design of the monochromator unit allows for a fixed beam position. The incident photon energy during the XAFS measurements around the Cl K-edge was varied between 2800 and 2950 eV in variable-energy steps, as described in the section before, using the Si(111) crystals for monochromatizing X-ray radiation originating from a bending magnet.

## Acknowledgements

This project has received funding from the European Unions Horizon 2020 research and innovation programme under the Marie Sklodowska Curie grant agreement no. 101029608. Furthermore, the European Commission funded this research within the H2020 ALION project under contract 646286 and the German Federal Ministry of Education and Research in the AlSiBat project under contract 03SF0486, and the project ALIBATT under contract 03XP0128E. The project 21GRD01 (OpMetBat) received funding from the European Partnership on Metrology, cofinanced by the European Union's Horizon Europe Research and Innovation Programme and by the Participating States.

## Conflict of Interest

The authors declare no conflict of interest.

## Data Availability Statement

The data that support the findings of this study are available from the corresponding author upon reasonable request.

## Keywords

electrochemically induced phase transitions, electronic structures, graphite intercalation compounds, superlattices, thermodynamics, X-ray absorption fine structure

Received: October 5, 2023
Revised: December 4, 2023
Published online: